\documentclass[referee]{raa}
\usepackage{graphicx,times,epsf}
\usepackage{natbib}
\newcommand{\ciii}{\ion{C}{iii}}
\newcommand{\civ}{\ion{C}{iv}}
\newcommand{\oiv}{\ion{O}{iv}}
\newcommand{\siiv}{\ion{Si}{iv}}
\newcommand{\neviii}{\ion{Ne}{viii}}

\begin{document}
\title{Transition-region explosive events produced by
plasmoid instability}

\volnopage{ {\bf 2018} Vol.\ {\bf X} No. {\bf XX}, 000--000}
   \setcounter{page}{1}

\author{Dong Li\inst{1,2}}
\institute{$^{1}$Key Laboratory of Dark Matter and Space Astronomy, Purple Mountain Observatory, CAS, Nanjing, 210034, China; {\it lidong@pmo.ac.cn}\\
          $^{2}$CAS Key Laboratory of Solar Activity, National Astronomical Observatories, Beijing 100012, China \\
%% Please give the E-mail address of the author, to whom future correspondence and offprint requests will be sent.
}

\date{Received ; accepted }

\abstract{Magnetic reconnection is thought to be a key process in
most of solar eruptions. Thanks to high-resolution observations and
simulations, the studied scale of reconnection process has become
smaller and smaller. Spectroscopic observations show that the
reconnection site can be very small, which always exhibits a bright
core and two extended wings with fast speeds, i.e.,
transition-region explosive events. In this paper, using the PLUTO
code, we perform a 2-D magnetohydrodynamic simulation to investigate
the small-scale reconnection in double current sheets. Based on our
simulation results, such as the line-of-sight velocity, number
density and plasma temperature, we can synthesize the line profile
of \siiv~1402.77~{\AA} which is a well known emission line to study
the transition-region explosive events on the Sun. The synthetic
line profile of \siiv~1402.77~{\AA} is complex with a bright core
and two broad wings which can extend to be nearly 200~km~s$^{-1}$.
Our simulation results suggest that the transition-region explosive
events on the Sun are produced by plasmoid instability during the
small-scale magnetic reconnection.  }

\authorrunning{Dong~Li}            %author_head in even pages
\titlerunning{Transition-region explosive events produced by plasmoid instability}  % title_head in odd pages
\maketitle

%________________________________________________ sections below
\section{Introduction}
\label{sect:intro}

The break and re-join of magnetic fields are generally thought to be
the process of magnetic reconnection. It is always believed as a
basic process of energy release on the Sun, and usually accompanied
by the particle acceleration. Almost all the solar
eruptions which are related with the magnetic energy
release are thought to be associated with magnetic reconnection,
i.e., coronal mass ejections \citep{Lin05,Guo13}, emerging flux
regions \citep{Zhao12,Yang14}, bipolar magnetic regions
\citep{Jiang14,Li17}, solar flares \citep{Petschek64,Liu10,Yan18},
solar filaments \citep{Shen15,Li16,Liy17}, bright points
\citep{Priest94,Li12,Zhao17}, and transition-region explosive events
\citep{Innes97,Ning04,Huang17}. That is to say, the Sun can provide
various observational features of magnetic reconnection in a wide
range of scales, i.e., from $\sim$1000~Mm to $\sim$1~Mm. The
diagnostics of magnetic reconnection often depend on the imaging
observations \citep{Masuda94,Tian14,Xue16}, the spectroscopic
observations \citep{Curdt11,Cheng15,Li18,Tian18}, and also the
magnetohydrodynamic (MHD) simulations
\citep{Jin96,Khomenko12,Yangl15}. Over the past few decades, many
observational features have been identified as the evidences of
magnetic reconnection on the Sun, such as the X-type structures
\citep{Aschwanden02,Su13,Yang15}, the magnetic null point
\citep{Sun12,Zhang12}, the current sheets \citep{Lin05,Liu10,Xue18},
the reconnection inflows or outflows
\citep{Liu13,Ning14,Sun15,Lid16}, the high-energetic particles
\citep{Li07,Li09,Klassen11}, and also the bi-directional jets
\citep{Dere91,Winebarger02,Innes13,Hong16}.

Transition-region explosive events are detected frequently in the
quiet or active Sun. They usually display as the broad non-Gaussian
profiles with a high velocity of about 100~$km~s^{-1}$ in the
emission lines, i.e., \siiv~1393~{\AA} \citep{Innes97,Ning04} and
1402.77~{\AA} \citep{Huang14,Innes15}, \civ~1548~{\AA} and
\oiv~1032~{\AA} \citep{Perez99,Innes13}. Usually, these emission
lines are formed between 6$\times$10$^4~K$ (e.g., \ciii) and
7$\times$10$^5~K$ (e.g., \neviii) \citep{Wilhelm07} in the
transition regions. The transition-region explosive events were
firstly detected by the spectra from High-Resolution Telescope and
Spectrograph (HRTS) observations \citep{Brueckner83}, and then
described in more detailed by \cite{Dere89} and \cite{Dere94}. Their
typical size is found to be about 1500~km, and the lifetime is
around 60~s. Spectroscopic observations
\citep[e.g.,][]{Dere89,Innes97,Ning04} further show that
they are always exhibiting a line asymmetry, i.e., the extended blue
or red wings, whose typical speed can be $\sim$100~$km~s^{-1}$. The
single event observed by a line profile often exhibits the blue and
red shifts simultaneously, which is regarded as the small-scale jet
with bi-directions in the transition region, and it is thought to be
associated with the small-scale magnetic reconnection on the Sun
\citep{Innes97}.

MHD numerical simulations have been applied to the magnetic
reconnection in the corona \citep{Shen11,Ni12}, the transition
region \citep[e.g.,][]{Innes99,Sarro99,Roussev01}, and the
chromosphere or even the photosphere \citep{Ni15,Ni16} on the Sun.
Previous simulations of the transition-region explosive events were
often based on the Petschek mechanism \citep{Petschek64}. The upward
and downward motions of magnetic islands were thought to be
associated with the blue and red shifts in the explosive events
\citep{Jin96}. Further simulation \citep{Roussev01} indicated that
the reconnection with an X-point in the transition region produced a
large blue shift ($\sim$100~$km~s^{-1}$), but a small red shift.
Meanwhile, \cite{Innes99} performed a compressible MHD simulations
of the small-scale explosive events based on the Petschek model, and
they found that the simulations could well reproduce the blue or red
shifts with high velocities, but failed to explain the bright core
near the line center with low velocities observed by the spectra
\cite[e.g.,][]{Dere91,Innes01}. Based on the large-scale MHD
simulations
\citep[e.g.,][]{Bhattacharjee09,Heggland09,Huang10,Huang17}, the
magnetic reconnection proceeds via the plasmoid instability has been
proposed. Now, this model has also been applied to explain the
small-scale explosive events which exhibit the brightening both at
spectral line core and two extended wings, i.e., the line profile of
\siiv\ 1402.77~{\AA} \citep[e.g.,][]{Innes15}.

In this paper, we perform a 2-D MHD simulation of magnetic
reconnection at a small scale which is produced in double current
sheets with the PLUTO code \citep{Mignone07,Mignone12}. We also
synthesize the line profiles which formed in the transition region,
i.e., \siiv\ 1402.77~{\AA} \citep{Tian17}. Our result suggests that
the transition-region explosive events on the Sun are the
small-scale reconnection sites, which is consistent with the IRIS
\citep{Dep14} observations \citep[see.,][]{Innes15}.

\section{Numerical method}
\subsection{MHD simulation}
In this paper, the PLUTO code \citep{Mignone07} with adaptive mesh
refinement (AMR) \citep{Mignone12} is applied to perform the MHD
simulation. Briefly, a Newtonian fluid with density ($\rho$),
velocity ($\mathbf{v}$) and magnetic field ($\mathbf{B}$) is
considered, then the forms of single-fluid MHD Equations are

\begin{eqnarray}
    \frac{\partial\rho}{\partial t} + \nabla\cdot(\rho\mathbf{v}) & = &  0, \nonumber \\
    \frac{\partial(\rho\mathbf{v})}{\partial t} + \nabla\cdot(\rho\mathbf{v}\mathbf{v} - \mathbf{B}\mathbf{B}) + \nabla p_{t} & = &  0, \\
    \frac{\partial E}{\partial t} + \nabla\cdot[(E + p_{t})\mathbf{v} - (\mathbf{v}\cdot\mathbf{B})\mathbf{B}] & = & -\eta\cdot\mathbf{J}\times\mathbf{B}-\Lambda, \nonumber \\
    \frac{\partial\mathbf{B}}{\partial t} - \nabla\times(\mathbf{v}\times\mathbf{B}) & = & -\nabla\times(\eta\mathbf{J}). \nonumber
 \label{mhd_e}
\end{eqnarray}

\noindent where $p_{t} = p + \mathbf{B}^{2}/2$ is the total pressure
which consists of thermal and magnetic pressures. $E = p/(\gamma -
1) + \frac{1}{2}\rho\mathbf{v}^{2}+\frac{1}{2}\mathbf{B}^{2}$ is the
total energy density at the ideal state, and $\gamma$ is the
specific heat ratio. $\mathbf{J} = \nabla\times\mathbf{B}$ is the
electric current density, and $\eta$ is the resistivity. Our
simulation considers the Ohmic heating and optically thin radiative
cooling (see., equation \ref{cool_e}), but the dissipative effects
caused by the viscous and gravity are ignored. Notice that only the
hydrogen gas is considered in our 2-D MHD model.

\begin{equation}
    \Lambda = n^{2}\tilde{\Lambda}(T), \quad\mbox{with}\quad
          n = \frac{\rho}{\mu m_u}.
 \label{cool_e}
\end{equation}

\begin{figure} %%%%%%%%%%%%%%%%%%
\centering
\includegraphics[width=\linewidth,clip=]{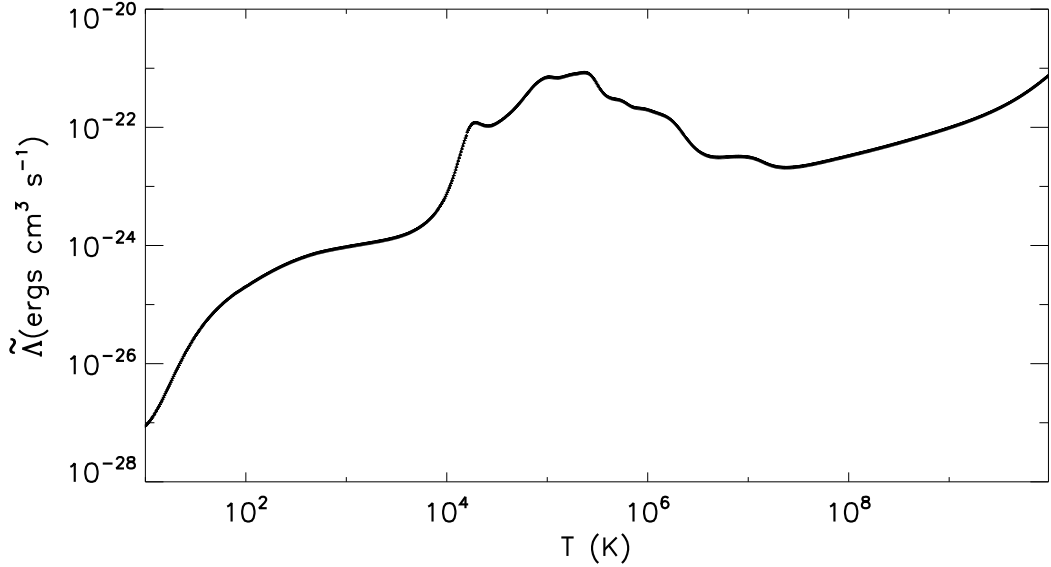}
\caption{Tabulated cooling rates in log$-$log space, the data is
derived from \cite{Mignone07}.} \label{cool}
\end{figure}

\noindent Here $n$ is the number density, $\mu$ is the mean
molecular weight, and $m_u$ is the atomic mass unit. In our model,
the cooling rates ($\tilde{\Lambda}$) are discrete as a table
sampled. Figure~\ref{cool} gives the cooling rates
\citep[see.,][]{Mignone07} used in this paper.

In our simulation, we consider a Cartesian 2-D uniform grid x, y
$\in$ [0, 2], with 64 $\times$ 64 zones at the coarse level, and
with periodic boundary conditions along all the two directions. To
trigger the magnetic reconnection, the initial magnetic field is
defined only in the vertical (Y) direction, as shown in
equation~\ref{magf_e} with $B = 1.0$ \citep[see.,][]{Mignone12}.

\begin{equation}
  B_{y} = \left\lbrace \begin{array}{l}
        -B ~~~~~~\quad\mbox{if}\quad |x - 1| \leq 0.5, \\
        ~~ B ~~~~~~\qquad\mbox{otherwise}.
        \end{array} \right.
\label{magf_e}
\end{equation}

\noindent In X direction, a small initial perturbation to the
velocity is set to start the reconnection, i.e., $v_{x} =
0.1\times\ v_0 sin(\pi\cdot y$). The initial thermal pressure is
given by the magnetic filed and plasma beta ($\beta$), such as $p =
\beta\cdot B^{2}/2$. The MHD fluid has an uniform density at the
initial beginning, which is $\rho = 1.0$. The temperate can be
obtained from the density and thermal pressure in the code, i.e., $T
= p/\rho$. In order to obtain the high-resolution grid and save the
computing time, the AMR is allowed for the MHD simulation. The
four-level refinement is activated for carrying out the integration
with the TVDLF solver. Finally, we perform the reconstruction on
characteristic variables rather than primitive values and set the
monotonized central difference limiter (MC\_LIM). In the whole
simulation, we assume an ideal gas state with the specific heat
ratio ($\gamma$) of 5/3, and the mean molecular weight ($\mu$) of
1/2.

Generally speaking, PLUTO is working with the non-dimensional (or
code) unit, which is described above. However, we have considered
the radiative cooling process, and the dimensional constants are
essential to scale data values to physical units. Thus, we have to
introduce the specific length and energy scale so that they can
compare with the dynamical advection scales. Therefore, we can
specify three fundamental units, they are the length ($L_0$), the
velocity ($v_0$), and the number density ($n_0$) which is
corresponding to the mass density of $\rho_0 = n_0 m_u$. The other
units can be derived from these three fundamental units, such as
time $t_0 = L_0/v_0$, pressure $p_0 = \rho_{0} v_{0}^{2}$, magnetic
field $B_0 = v_{0} \sqrt{4\pi \rho_{0}}$, and temperature $T_0 = \mu
m_u v_{0}^{2}/k_B$, where $k_B$ is the Boltzmann constant. Noting
that all these dimensional units are c.g.s, i.e., $cm$, $s$, and
$gr$. During the entire process of MHD simulation, an uniform
resistivity ($\eta$) is employed.

Table~\ref{t_f} lists the characteristic and initial values of these fundamental
units in our 2-D simulation. The characteristic length is set to be
10$^8~cm$, and thus the simulation box has a size of 2$\times$2~Mm.
The characteristic number density and velocity are set to be
10$^{10}~cm^{-3}$ and 2$\times$10$^7~cm~s^{-1}$ at first,
respectively, which leads to a characteristic plasma beta of $\beta
\approx 1.67$, as shown in the second lines (CPs) of table~\ref{t_f}. Therefore, the MHD flows have an initial uniform
density of 10$^{10}~cm^{-3}$, the initial magnetic field of about
10~$G$, and a small initial perturbation velocity of
2$\times$10$^6~cm~s^{-1}$, this gives an initial plasma beta of $\beta
\approx 0.02$, as indicated in the third lines (IPs) of table~\ref{t_f}. Finally, an uniform resistivity ($\eta$ = 10$^{-5}$)
is performed.

\begin{table}
\caption{The characteristic parameters (CPs) and the initial parameters (IPs) in our 2-D MHD simulations.}
\centering \setlength{\tabcolsep}{4pt}
\begin{tabular}{c c c c c c c c}
 \hline\hline
        & $L_0$ (cm) & $n_0$ (cm$^{-3}$)  &  $v_0$ (cm~s$^{-1}$)  & $T_0$ (K)          & $B_0$ (G) &  $\beta$  &  $\eta$   \\
  \hline
CPs  & 10$^8$    &  10$^{10}$          &   2$\times$10$^7$     & 2.4$\times$10$^6$  & $\sim$10   &  1.67   & 10$^{-5}$    \\
 \hline
IPs & 10$^8$    &  10$^{10}$          &   2$\times$10$^6$     & 2.4$\times$10$^4$  & $\sim$10   &  0.02   & 10$^{-5}$    \\
\hline \hline
\end{tabular}
\label{t_f}
\end{table}

\subsection{Line synthesis}
The values of six variables, such as the number density($n$), the
total energy (E), the velocities ($v_x$, $v_y$) and the magnetic
fields (B$_x$, B$_y$) are derived at each time step by solving the
2-D MHD equations, then the plasma temperature ($T$) can be derived
by using the values of these variables. Using these variables, and
assuming that the emission line is made up with a group of Gaussian
profiles, we can synthesize the line profile with
equation~\ref{line_e} \citep[see also.,][]{Hansteen10,Yuan16}, i.e.,
\siiv\ 1402.77~{\AA}, which is a well known spectral line to study
the transition-region explosive events from IRIS observations
\citep[e.g.,][]{Innes15}.

\begin{equation}
    I(v) = \frac{\Sigma}{4\pi R_{AU}^2}\int{\phi(v_s)~n_e~n~g(T)}ds.
 \label{line_e}
\end{equation}

\noindent Here the integral intensity of spectral line is carried
out along the chosen light-of-sight (LOS). $n_e$ is the electron
number density, and assuming that $n_e = n$, $\Sigma$ is the area of
the MHD simulation region, $R_{AU}$ is the distance between the Sun
and the Earth, $\frac{\Sigma}{4\pi R_{AU}^2}$ represent the ratio
between the spectral line intensity which radiate from the Sun and
that we have received on the Earth \cite[e.g.,][]{Winebarger99}.
$g(T)$ is the contribution function of the specified emission line,
i.e., \siiv~1402.77~{\AA}, which can be obtained from the CHIANTI
database version 8.0 \citep{Del15}. $\phi(v_s)$ is a Gaussian
function with a variable of $v_s$ (the unit of velocity is
$km~s^{-1}$ here), see., equation~\ref{gass_e}.

\begin{equation}
    \phi(v_s) = \frac{1}{\sqrt{\pi}\Delta v_D}e^{-\frac{(v-v_s)^2}{2\Delta v_D^2}}, \quad\mbox{with}\quad
    \Delta v_D = \sqrt{\frac{2k_BT}{m}}.
 \label{gass_e}
\end{equation}
Where $\Delta v_D$ represents the thermal broadening from the
simulation temperature $T$, and $m$ is the mass of the radiating
ion, while $v_s$ is the fluid speed projected along the LOS
direction, i.e., the velocity in y-direction ($v_y$) from
the numerical results, as shown in Figure~\ref{svtn}.

\section{Results}
\subsection{MHD simulation results}
\begin{figure} %%%%%%%%%%%%%%%%%%
\centering
\includegraphics[width=\linewidth,clip=]{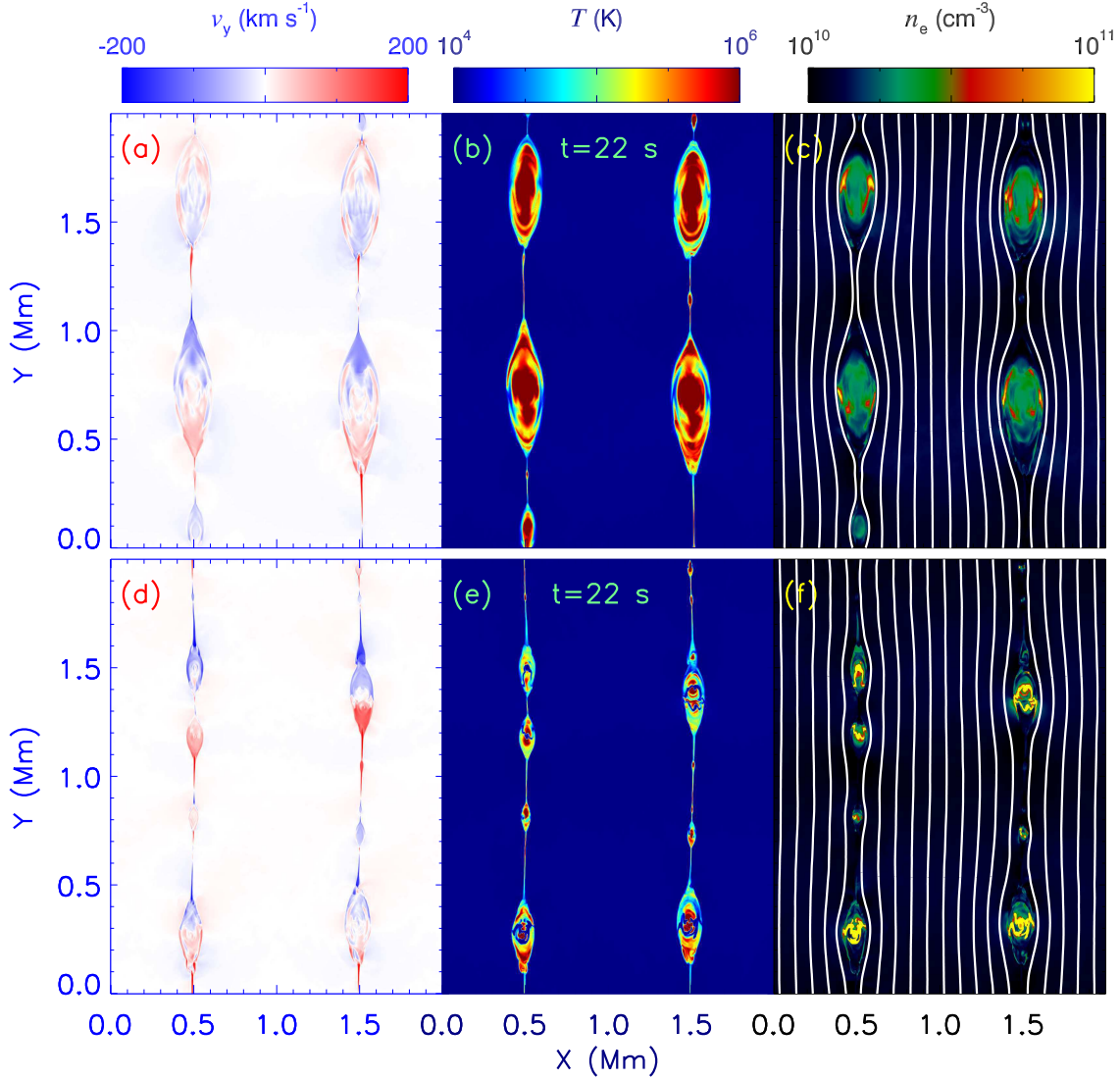}
\caption{The MHD simulation results without (upper) and with
(bottom) cooling process at t = 22~s: velocity (a, d), plasma
temperature (b, e), and number density (c, f). The white profiles
represent the magnetic field lines.}\label{svtn}
\end{figure}

Figure~\ref{svtn} shows the variables which derived from our 2-D MHD
simulation at t = 22~s. The upper panels gives the simulation
variables without radiative cooling process, such as the velocities
along Y direction ($v_y$), the plasma temperature ($T$), and the
number density ($n_e$). All these simulation variables show that
every magnetic island includes an O-point in its center and a pair
of X-point at its sides. These magnetic islands in the
double current sheets interact with each other. The Joule heating
and shocks inside these islands are the possible mechanisms to heat
the plasmas \citep{Ni15,Ni16}, and could further produce the bright
core and extended enhanced wings \citep{Innes15}. The velocities at
two sides of magnetic islands can be reached a fast speed of around
200~$km~s^{-1}$, whatever the blue or red shifts, while the
velocities in the centers of magnetic islands is slow, as shown in
panel~(a). However, the plasma temperature is very high in the
double current sheets, especially in the magnetic islands. The
plasma temperature in the center and two sides of magnetic islands
can be as high as 10$^6~K$ (panel~b), which is much higher than the
formation temperature of \siiv\ 1402.77~{\AA}, i.e.,
8$\times$10$^4~K$. Therefore, we expect the radiative cooling
process to decrease the temperature inside the reconnection region,
as seen $\Lambda$ in equation~1.

The cooling function is defined by the number density ($n$) and
cooling rates (see., equation~\ref{cool_e}), while the cooling rates
($\tilde{\Lambda}$) are from the table which given by
\cite{Mignone07}. Figure~\ref{cool} shows the cooling rates depend
on the temperature, which clearly shows that the cooling rates are
strongly sensitive to the temperature between about 10$^{4}~K$ and
10$^6~K$. This is just what we are interested, as the
formation temperature of \siiv\ 1402.77~{\AA} is exactly between
them. In fact, the cooling rates are also sensitive at much higher
temperature, i.e., $> 10^8~K$, but it is out of scope of this study.

The bottom panels in Figure~\ref{svtn} show the MHD simulation
results with the radiative cooling process at t = 22~s.
Similar as the upper panels in Figure~\ref{svtn}, multiple
magnetic islands with O-points in their centers and X-points at
their two sides appear, they are interact with each other. The
velocities at two sides of magnetic islands can be around
200~$km~s^{-1}$, while the velocities in the centers of magnetic
islands is slow, as shown in panel~(d). Moreover, the plasma
temperature in the reconnection region are decreasing, but the
number density is increasing at the same time. The plasma
temperature in the double current sheets is cooling down to form the
\siiv\ 1402.77~{\AA} line, i.e., in the centers and two sides of the
magnetic islands.

\subsection{Synthesis line profiles}
\begin{figure} %%%%%%%%%%%%%%%%%%
\centering
\includegraphics[width=\linewidth,clip=]{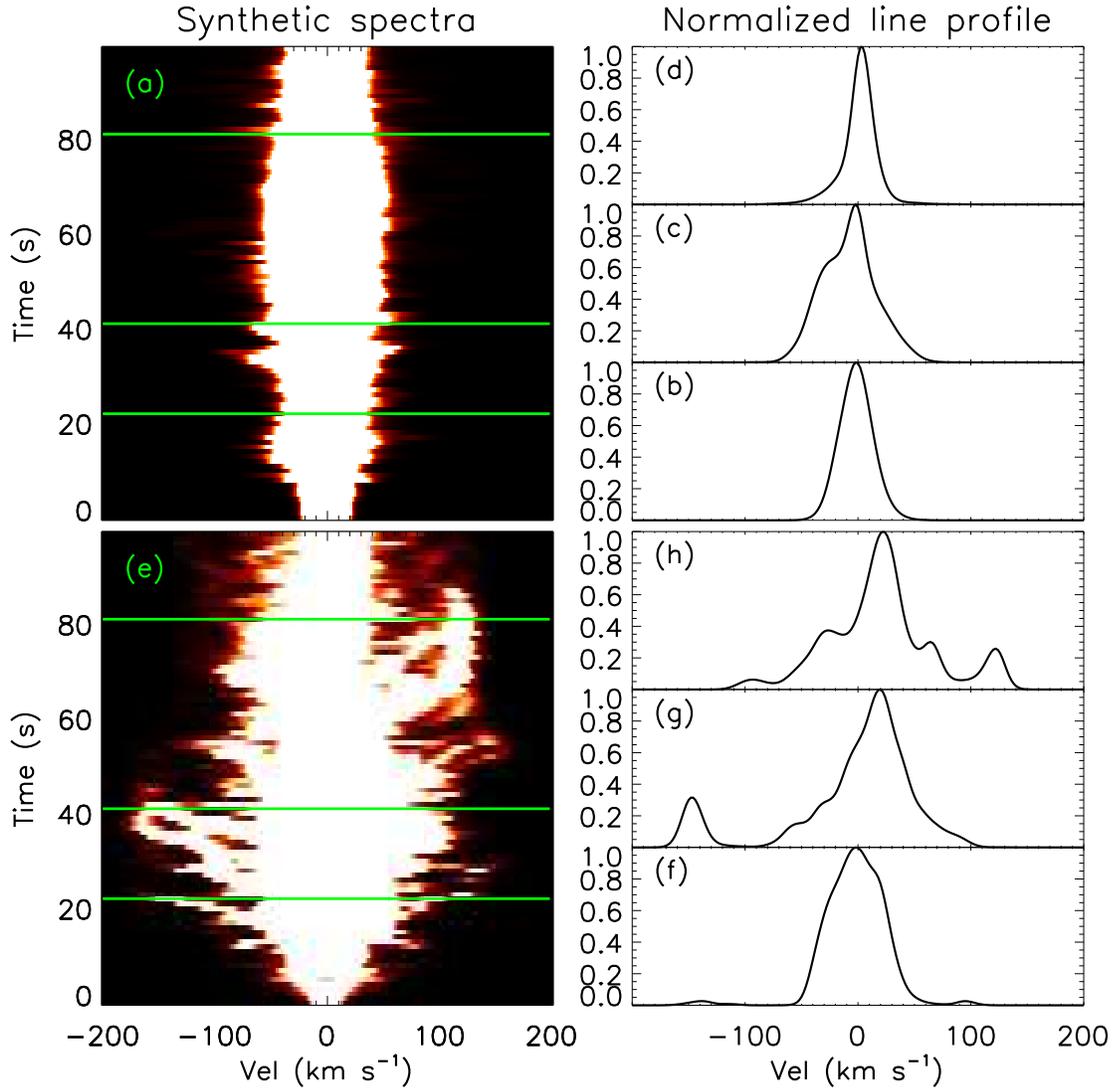}
\caption{Line profiles obtained from the 2-D MHD simulations. Left:
time evolution of the line profiles of \siiv\ 1402.77~{\AA} without
(a) and with (e) the cooling process. Right: Line profiles indicated
by the green horizontal lines on the left panel.} \label{line}
\end{figure}

Using equation~\ref{line_e} and the simulation variables, we can
synthesize the line profiles of \siiv~1402.77~{\AA}, as shown in
Figure~\ref{line}. Here, the velocity distribution (Vel) is
the velocity (v) on the left side of equation (4), and the final
value of the intensity for each given v (Vel) as shown in
Figure~\ref{line} is then calculated by adding all the integral
intensity I(v) at each X point. Panel (e) shows the
evolution of the spectral line with time in the case with radiative
cooling. At the beginning of t~$\approx$ 0$-$5~s, only the line
core is bright, and it is also very narrow. Then the line core
becomes broader and broader with time, such as at t~$\approx$
5$-$15~s. Next, two extended line wings appear to brighten and move
to blue and red wings simultaneously, i.e., t~$\approx$ 15$-$85~s,
the maximum speed can reach up to nearly 200~$km~s^{-1}$, as shown
in panel~(g) and (h). Finally, both blue and red wings of the line
profiles disappear, leaving only the bright and narrow core, i.e.,
t~$\approx$ 85$-$100~s. During the time intervals between t
$\approx$ 15$-$85~s, the spectral line is always non-Gaussian
profile and asymmetrical, and the line core is brighter than its two
extended wings, while the blue and red wings become brighter and
brighter from the beginning of the explosive events, but disappear
at the end, see the panels of (g)$-$(h). This evolution of line
profiles is very similar to that of the transition-region explosive
events observed by IRIS \citep[see.,][]{Innes15}. On the other hand,
the duration of this time interval is about 70~s, which is also
similar to the lifetime of the transition-region explosive events in
the spectroscopic observations
\citep[e.g.,][]{Dere89,Dere94,Ning04,Innes15}.

To compare the simulation results with and without cooling process,
we also calculate the spectral line in the case without
radiative cooling by using equation~\ref{line_e}, as shown in the
upper panels in Figure~\ref{line}. Panel~(a) gives the evolution of
the spectral line with time, and panels~(b)$-$(d) display the line
profiles at three times. All these panels show that the spectral
line is symmetrical and Gaussian profile in most of the simulation
times. The line core is always bright and broad, but the two
extended wings do not appear during our simulation times. We also
notice that the line profile in panel~(c) exhibits line
asymmetrical, but it could not identified as the transition-region
explosive event, because it only brightens in the line core.

In our 2-D simulation, the small-scale magnetic reconnection
produces the O-point centers with a slow speed and the opposite
directional jets at two X-point sides with a fast speed, as
indicated by the bright core and blue/red shifts in the line
profiles. At the same time, the double current sheets become
unstable and break up into several small plasmoids, the centers of
these small plasmoids move with different velocities, and they are
the broad and bright core in the line profiles. While two sides of
these small plasmoids move toward to the opposite directions with a
fast speed, and they are the blue/red shifts in the line profiles,
as seen in Figure~\ref{line}. The blue and red shifts of the line
profiles can reach nearly 200~$km~s^{-1}$, while the bright core of
the line profiles only exhibits a slow velocity, and becomes broader
and broader with time. Our simulation results are well consistent
with the spectroscopic observations from IRIS
\citep[e.g.,][]{Innes15}.

\subsection{The cooling process}
The key point of our 2-D MHD simulation is the cooling process,
since we aim to simulate the transition-region explosive event in
\siiv~1402.77~{\AA} that are characterized by a fast velocity but a
low temperature. As shown in the upper panels of Figure 2, although
the velocity at the two sides of the magnetic islands is nearly
200~km~s$^{-1}$, the plasma temperature at these regions is very
high, and the highest temperature could be nearly 10$^6$~K.
Moreover, the plasma temperature in the centers of the magnetic
islands is also as high as $\sim$10$^{6}$~K. Such high temperature
is far more than the formation temperature of \siiv\ 1402.77~{\AA},
i.e., $\sim$8$\times$10$^4~K$. Therefore, in order to synthesize the
\siiv\ 1402.77~{\AA} line profiles, we introduce the cooling
function ($\Lambda$) in the energy conservation equation (see.,
equations~1). Thus it is possible to synthesize the line profile of
\siiv\ 1402.77~{\AA} under the condition of fast velocity and low
temperature. Figure~\ref{svtn} shows the MHD simulation
results at t = 22~s without (upper) and with (bottom) cooling
functions, respectively, which indicates that the cooling function
is working at the early stage of the explosive event.

\begin{figure} %%%%%%%%%%%%%%%%%%
\centering
\includegraphics[width=\linewidth,clip=]{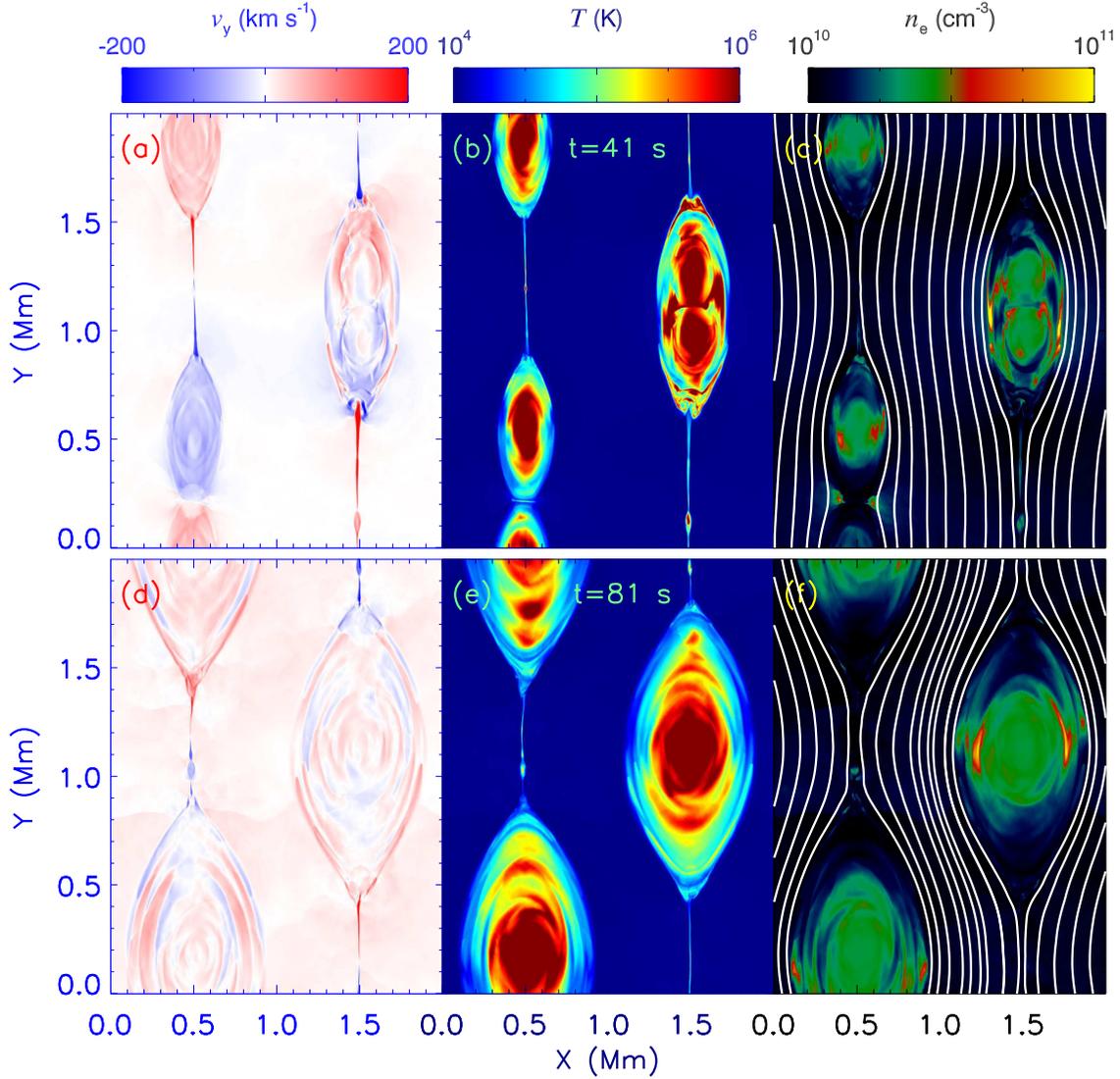}
\caption{Similar as the upper panels of Figure~\ref{svtn}, but at t
= 41~s and t = 81~s, respectively.} \label{lvtn0}
\end{figure}

\begin{figure} %%%%%%%%%%%%%%%%%%
\centering
\includegraphics[width=\linewidth,clip=]{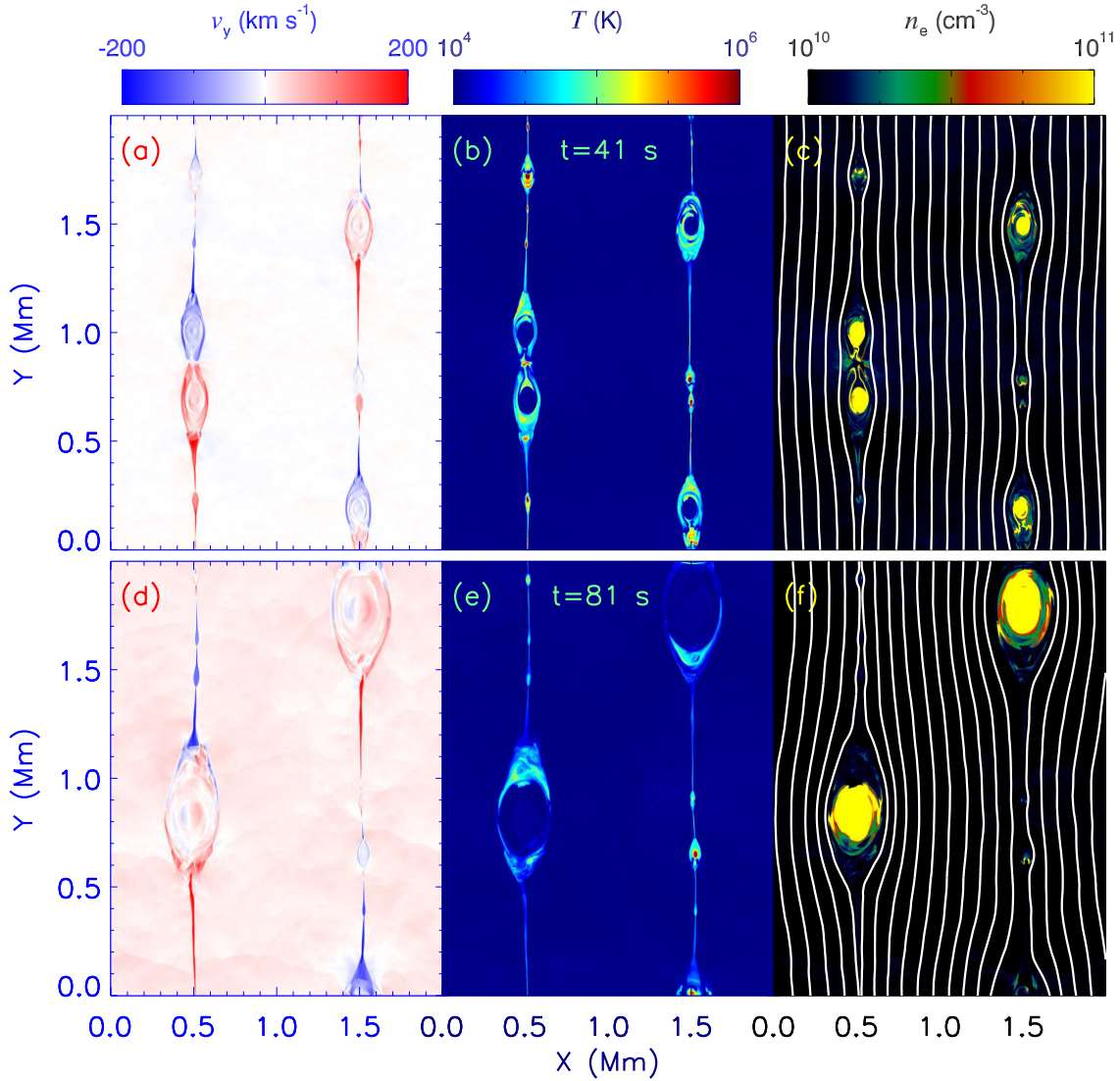}
\caption{Similar as the bottom panels of Figure~\ref{svtn}, but at t
= 41~s and t = 81~s, respectively.} \label{lvtn1}
\end{figure}

Figures~\ref{lvtn0} and \ref{lvtn1} further display our MHD
simulation results without and with the cooling process at the
middle and end time of the explosive event, such as t = 41~s and t =
81~s, respectively. Their line profiles are shown in
Figure~\ref{line}~(c)$-$(d) and (g)$-$(h), respectively. The
simulation results show that the double current sheets become
broader and broader with time if there is no cooling process, and
the plasma temperature in the reconnection regions are too hot to
produce the line profile of \siiv\ 1402.77~{\AA}, including the
center and two sides of the magnetic islands. However, when we
introduce the cooling function, the plasma temperature is cooling
down, no matter the centers or two sides of the magnetic islands.
Therefore, we can synthesize the \siiv~1402.77~{\AA} line profiles
from the 2-D MHD simulation results. Figure~\ref{line} (bottom)
shows the synthetic line profiles in \siiv~1402.77~{\AA}. It
exhibits a broad non-gaussian profile (f, g and h) with a bright
core and two extended (blue/red) wings (g and h).

\section{Conclusions and Discussions}
Using the PLUTO code, we perform a 2-D MHD simulation of the
small-scale magnetic reconnection in double current sheets. Then
using the simulation variables, such as the LOS velocity, number
density and plasma temperature, we synthesize the line profiles of
\siiv~1402.77~{\AA}. The spectral line shows a broad non-Gaussian
profile, it is complex with a bright core and two asymmetrical
extended wings (blue or red shifts with velocity of nearly
200~$km~s^{-1}$), and this process can last for $\sim$70~s, as shown
in Figure~\ref{line}. All these line features agree well with the
transition-region explosive events observed by IRIS
\citep[see.,][]{Innes15}. Their observations show that the spectral
line profiles of \siiv~1402.77~{\AA} at small-scale acceleration
sites are broad and asymmetrical, they have bright central cores and
extended blue/red wings with low intensity and fast velocity over
200~$km~s^{-1}$. The line emissions from the bright core and two
extended wings are spatially coincident and neither move
significantly during the lifetime of the explosive events. These
observational line features can also be found in our simulation, as
seen in the bottom panels Figure~\ref{line}. Here, the
double current sheet configuration is used to allow the periodic
boundaries. As we know, the boundaries around the transition-region
explosive events in the solar atmosphere is definitely not periodic.
Thus, the periodic boundary conditions make the plasmas to flow out
from one boundary and flow in from the other boundary in y-direction
in this work, which has affected the reconnection region and the
distributions of the plasma temperature and density. Therefore, the
synthesized spectral lines will possibly be different if the other
more realistic boundary conditions are applied. And we will check
the simulations by using an open boundary in the future work.

The explosive evens with bi-directional jets observed in the
transition regions are thought to be produced by the small-scale
magnetic reconnection on the Sun \citep{Innes97}. They have been
studied by many authors based on the spectroscopic observations
\citep[e.g.,][]{Dere94,Innes97,Ning04,Innes13,Huang14} and MHD
simulations \citep{Jin96,Innes99,Sarro99,Roussev01}. However, there
is always a contradiction between the observations and the
simulations of these line profiles. For example, spectroscopic
observations find that the line profiles are complex with both
bright cores and faint asymmetrical wings
\citep{Dere94,Innes97,Ning04,Innes15}, while the MHD
simulation results based on the Petschek model\citep{Petschek64} can
only reproduce the extended wing brightening
\citep{Innes99,Biskamp00,Roussev01}. This is because that the
Petschek mechanism only accounts for the line wings with fast
velocity, but it fails to reproduce the bright core with slow
velocity since there are not enough low-speed plasmas at the
diffusion region \citep[e.g.,][]{Innes99}. Recent large-scale MHD
simulations show that the plasmas with fast speeds may reproduce the
jets, while the plasmas with slow speeds are outside of the
diffusion region \citep{Heggland09,Ding11}. Inspiring from that,
\cite{Innes15} present that the bright core of line profile is from
the heating of the background plasmas. That is to say, the emissions
from jets and backgrounds are spatially off-set, and this can result
into the spatial off-sets between the line core and two extended
wings brightening. This reconnection model can explain the bright
core and extended wings very well. Our 2-D MHD simulations of
magnetic reconnection show a rapid growth of the magnetic islands
along the double current sheets. These magnetic islands are divided
by the fast jets and can well explain the bright core and the
extended wings of the line profiles in the explosive events.
Therefore, we conclude that the transition-region explosive events
with a bright core and two extended wings are most likely produced
by the plasmoid instability rather than the Petschek mechanism
\citep{Petschek64} during the small-scale magnetic reconnection on
the Sun. The bright emission of line core is contributed from the
high-density and low-velocity magnetic islands, as shown in
Figures~\ref{svtn} (bottom) and \ref{lvtn1}.

As mentioned in Section 3.3, the key point of our 2-D MHD simulation
is the cooling process. In this paper, the optically thin radiative
cooling is considered, which is dependent on the cooling rates and
the number density (see., equation~\ref{cool_e}). The tabulated
cooling rates (see., Figure~\ref{cool}) are used to decrease the
plasma temperature inside the reconnection region, and they seem to
work well (e.g., Figures~\ref{lvtn0} and \ref{lvtn1}), especially at
the temperature region between 10$^4~K$ and 10$^6~K$ \citep[see
also.,][]{Schmutzler93,Tesileanu08}. The thermal conduction is
neglected in our 2-D MHD simulation. We do not know if the
heat conduction will play a role on our 2-D simulation, and it will
be checked in the future work. On the other hand, we perform an
initial uniform density of 10$^{10}$~cm$^{-3}$ in the simulation,
which can be resulted into a maximum density of
$\sim$$10^{11}$~cm$^{-3}$. This value is consistent with the number
density in the transition region on the Sun. Therefore, our 2-D MHD
simulation conditions are very close to the situations in the solar
transition region, and our simulation results agree well with the
IRIS spectroscopic observations \citep[see.,][]{Innes15}. In
fact, \cite{Innes15} have already synthesized the line profile of
\siiv~1402.77~{\AA} by using the 2-D simulation results of magnetic
reconnection with plasmoid instabilities, the radiative cooling is
not included in their simulations. We add the radiative cool
process in our 2-D MHD simulations and obtain the similar results.
In the further, we wish to work together with them using their
method and code.

\cite{Innes15} firstly successfully apply the reconnection
process with plasmoid instability to explain the \siiv\ line profile
in the transition region events. The characteristic parameters and
initial conditions in their 2-D simulation have been described,
i.e., the  plasma density, magnetic field and
temperature in the inflow regions are set to be 10$^{10}$~cm$^{-3}$, 12~G, and
2$\times$10$^5$~K. In our 2-D simulations, the initial plasma density, magnetic
field and temperature  are
10$^{10}$~cm$^{-3}$, 10~G, and 2.4$\times$10$^4$~K, respectively.
Therefore, the initial plasma beta ($\beta$=0.02) in this work is
about five times smaller than that ($\beta$=0.1) in the inflow
region of \cite{Innes15}. In \cite{Innes15}, Harris Sheet is applied
as the initial configurations of magnetic field \citep{Guo14}, which
makes the initial plasma density and pressure to be non-uniform
across the current sheet, the initial plasma density and plasma beta
in the center of the current sheet is higher than those in the
inflow region. As shown in equation (3) in our simulation, the
absolute value of the initial magnetic field is uniform across the
current sheet, which makes the initial plasma density and pressure
also to be uniform. Therefore, the different initial plasma beta
($\beta$) and initial configurations of magnetic field, plasma
density and pressure are the possible reasons to cause the huge
difference on the plasma heating between our work and
\cite{Innes15}. The plasma in the simulation work of \cite{Innes15}
is not strongly heated as shown in our work, the maximum plasma
temperature in \cite{Innes15} only reaches the characteristic value
2$\times$10$^5$~K. Therefore, the \siiv\ line profile can be
synthesized well by using their 2-D simulation results without
radiative cooling. Therefore, the radiative cooling is possibly not
important in some of the transition region reconnection events.

In our 2-D simulation, the magnetic diffusion is set as a constant,
and the highest resolution is (64$\times$2$^4$)*(64$\times$2$^4$).
As we know, The resolution must be high enough in the simulations,
otherwise the numerical diffusion will be higher than the magnetic
diffusion set in the MHD equations and the numerical diffusion
instead of the magnetic diffusion dissipate the magnetic energy.
Therefore, a higher refinement level is used in our 2-D simulation,
such as (64$\times$2$^5$)*(64$\times$2$^5$). And the simulation
results by using a higher refinement level are very similar as those
in our above work, i.e., the plasma temperatures both in the centers
and at two sides of magnetic islands are cooling down. Therefore,
the resolution in our work is high enough in our work.

\begin{acknowledgements}
We acknowledged the anonymous referee for his/her inspiring
suggestions and constructive comments. The author would like to
thank Dr.~D.~E.~Innes for her valuable suggestions to complete this
paper. PLUTO is a freely-distributed software for the numerical
solution of mixed hyperbolic/parabolic systems of partial
differential equations (conservation laws) targeting high Mach
number flows in astrophysical fluid dynamics. This study is
supported by NSFC under grants 11603077, 11573072, 11790302,
11333009, the CRP (KLSA201708), the Youth Fund of Jiangsu Nos.
BK20161095, and BK20171108, and the National Natural Science Foundation of China (U1731241), 
the Strategic Priority Research Program on Space Science, CAS (nos. XDA15052200 and XDA15320301).
The Laboratory NO. 2010DP173032.
\end{acknowledgements}

\label{lastpage}
\end{document}